\documentclass[preprint2]{aastex}

\def\kms{\mbox{km/s}}

\def\kpch{\mbox{$h^{-1}$kpc}}

\def\LCDM{{$\Lambda$CDM}}

\def\mpch{\mbox{$h^{-1}$Mpc}}

\def\msun{\mbox{M$_\odot$}}
\def\msunh{\mbox{$h^{-1}$M$_\odot$}}
\def\Mvir{\mbox{$M_{\rm vir}$}}
\def\nfw{\mbox{{\tiny NFW}}}
\def\ome{\mbox{$\Omega_0$}}
\def\omel{\mbox{$\Omega_\Lambda$}}

\def\pch{\mbox{$h^{-1}$pc}}

\def\rvir{\mbox{$r_{\rm vir}$}}
\def\sige{\mbox{$\sigma_8$}}
\def\Vmax{\mbox{$v_{\rm max}$}}
\def\Vc{\mbox{$V_c$}}
\def\ltsima{$\; \buildrel < \over \sim \;$}
\def\lsim{\lower.5ex\hbox{\ltsima}}
\def\gtsima{$\; \buildrel > \over \sim \;$}
\def\gsim{\lower.5ex\hbox{\gtsima}}
\def\mathnew{\mathsurround=0pt}
\def\simov#1#2{\lower .5pt\vbox{\baselineskip0pt
    \lineskip-.5pt\ialign{$\mathnew#1\hfil##\hfil$\crcr#2\crcr\sim\crcr}}}

\def\'#1{\ifx#1i{\accent"13\i}\else{\accent"13#1}\fi}
\def\eg{e.g.,}

\begin{document}
\title{Dwarf Dark Matter Halos}
\author{P. Col\'in}
\affil{Instituto de Astronom\'ia, Universidad Nacional Aut\'onoma
de M\'exico, C.P. 04510, M\'exico, D.F., M\'exico}
\author{ A. Klypin, O. Valenzuela}
\affil{Astronomy Department, New Mexico State University, Box 30001, Department
4500, Las Cruces, NM 88003-0001, U.S.A.}
\author{Stefan Gottl\"ober}
\affil{Astrophysikalisches Institut Potsdam,
An der Sternwarte 16,
14482 Potsdam, Germany}
%{\em draft, {\today}}}

\begin{abstract}
We study properties of dark matter halos at high redshifts $z=2-10$
for a vast range of masses with the emphasis on dwarf halos with
masses $10^7-10^9 \msunh$.  We find that the density profiles of {\it
relaxed} dwarf halos are well fitted by the NFW profile and {\it do
not have cores}.  We compute the halo mass function and the halo spin
parameter distribution and find that the former is very well
reproduced by the Sheth \& Tormen model while the latter is well
fitted by a lognormal distribution with $\lambda_0 = 0.042$ and
$\sigma_{\lambda} = 0.63$.  
We estimate the distribution
of concentrations for halos in mass range that covers six orders of
magnitude from $10^7$ \msunh\ to $10^{13}$ \msunh, and find that the
data are well reproduced by the model of Bullock et al. The
extrapolation of our results to $z = 0$ predicts that present-day
isolated dwarf halos should have a very large median concentration of
$\sim 35$.  We measure the subhalo circular velocity functions for
halos with masses that range from $4.6 \times 10^9$\msunh\ to
$10^{13}$ \msunh\ and find that they are similar when normalized to
the circular velocity of the parent halo. Dwarf halos studied in this
paper are many orders of magnitude smaller than well-studied cluster-
and Milky Way-sized halos. Yet, in all respects the dwarfs are just
down-scaled versions of the large halos. They are cuspy and, as
expected, more concentrated. They have the same spin parameter
distribution and follow the same mass function that was measured for
large halos.

\end{abstract}


\section{Introduction}

Dwarf galaxies with virial masses $10^7-10^{9}\msun$ are the smallest
virialized objects in the universe that show evidence of dark matter
\citep[][and references therein]{mateo98}.  Some of these galaxies are
so small that they can be detected only in the Local Group.  Large
amounts of  dark matter along with the proximity to the Milky Way
make  dwarf galaxies an ideal laboratory for testing the
hierarchical cosmologies such as the standard $\Lambda$CDM. The
rotation curves of dwarf irregular galaxies and the velocity dispersions
of dwarf spheroidal galaxies have been used in the last decade to
constrain the density of the dark matter in the central parts of these
galaxies and, thus, to constrain the structure of dark matter (DM)
halos that host the galaxies.  So far the results indicate that DM
halos should have flat cores \citep{CF1988, CB1989, FP1994, Moore94,
Burkert1995}. This is inconsistent with the cold dark matter theory,
which predicts cuspy cores with $\rho \propto r^{-1}$
\citep[][hereafter NFW]{DC1991, NFW95,NFW96, NFW97} or $\rho \propto
r^{-1.5}$ \citep{Moore1999a}.  The controversy of the flat versus
cuspy DM cores is still not resolved.  Recent observations and
analysis of dwarf and low surface brightness (LSB) galaxies rotation
curves continue to suggest a dark matter halo density profile with a
relatively flat core \citep{deBlok, WdBW2003}, but see also
\citet{Swaters2003, Rhee2003}.
 
A second somewhat related problem faced by the CDM cosmogony is the
excessive substructure predicted by this model in Milky Way-sized
halos as compared with what is observed \citep{Klypin1999b,
Moore1999b}.  Within the CDM framework this problem finds a natural
explanation in the reionization of the universe: satellites form only
inside subhalos that collapse early, before the universe was fully
ionized \citep{BKW2000, Somerville2002, Benson2002}. There may be a
different reading of the substructure problem which depends on how
peak circular velocities, \Vmax, are assigned to dwarf spheroidal
(dSph) galaxies (these galaxies comprise most of the Milky Way
satellites).  If these galaxies have a larger \Vmax\ than what was
previously expected, there is no disagreement: the small number of
satellites would be explained by the relatively scarcity by which
large subhalos are formed in the CDM model.  In this case every large
DM satellite hosts a dwarf galaxy. The absence of a stellar component
in the numerous small DM satellites could be explained by feedback
and/or reionization.  The problem seems to be degenerate: observed
dwarf galaxies can be hosted by either a relatively low-concentration
but high-\Vmax\ halo \citep{Hayashi2003, Stoehr02} or by a
high-concentration and low-\Vmax\ halo \citep{lokas02}.  Thus, the
concentration of dwarf DM halos is quite important.  Accurate
measurements of the concentration for a large sample of dwarf DM halos
is one of the goals of this paper.
 
%%AAK rephrased this paragraph
Attempts have been made to simulate dwarf halos with high
resolution within a cosmological context. \citet{Moore2001} simulated
one ``Draco''-sized halo ($\Mvir \sim 10^8 \msunh$) with several
million particles and with high force resolution.  Their fit of the
density profile favor the inner slope $-1.3$.  \citet{Ricotti} presented
several simulations including
%%AAK  $256^3$ particles -- that may confuse people. What is
%%                  important is the number of particles in a halo,
%%                  not the total. 
 one with a small (1 \mpch) box.
%%AAK  and had enough mass resolution to  -- that actually is not true: not enough
%%    study the structure of dwarf halos.
 The dwarf halos in this small box are resolved with only few tens of
thousands of particles at high redshifts.  \citet{Ricotti} finds that
at $z \sim 10$ the density profiles have inner slopes in the range
$[-0.4,-0.5]$.  Recently, \citet{Cen04} reproduced results of
\citet{Ricotti} using halos with very small number (few thousands) of
particles.  Cusps in these simulations are much flatter than that
quoted by Moore et al.  \citet{Navarro03} made high resolution
simulations of dwarf halos, but those were presented only at
$z=0$. Dwarf halos were found to have steep profiles. Thus, the issue
of the evolution of the cusps of dwarfs is still not settled.
Numerical effects are of special concern for low resolution
simulations. Yet, there are other issues, which require attension
including treatment of non-equilibrium features in the density
distribution.

 In this paper, we
also study cusps of dwarf halos.  But cusps is only one of the
aspects, which we are interested in.  The statistical properties of
this population of halos such as the mass function or the spin
parameter distribution are also studied.

\begin{deluxetable}{ccrcc}
\tablecolumns{5}
\tablewidth{0pc}
\tablecaption{Parameters of simulations}
\label{tab:simu}
\tablehead{\colhead{Box}  & \colhead{Mass resolution} &
\colhead{Force resolution} & \colhead{\sige} & \colhead{Name} \\
(\mpch) &   (\msunh) & (comoving \pch) &  & }
\startdata
  1 & $4.9 \times 10^3$  &  61  & 0.75 & A \\
 25 & $1.2 \times 10^6$  & 191  & 0.90 & B \\
 80 & $4.0 \times 10^7$  & 610  & 0.90 & C \\
 60 & $1.1 \times 10^9$  & 2000 & 1.00 & D \\
\enddata
\end{deluxetable}

Dwarf halos with masses $\Mvir \sim 10^7 - 10^9 \msunh$ studied in
this paper form at very high redshifts. If a halo of this low mass
remains isolated, it is expected to grow only 2-3 times since redshift
$z\approx 3$ until $z=0$ \citep{Bosch2002} with most of the mass
ending up in the  outer part of the halo. Thus, the halo
profile should not change much since $z=3$.  The dwarfs may be
accreted  by larger halos and become satellites. If that happens,
they are significantly stripped and lose most of their mass in
outer regions. In any case, it is important to know what was the
structure of the dwarfs at high redshifts. This is why we focus on the
structure of the dwarfs at $z\approx 3$.

The paper is organized as follows. In \S 2 we present the cosmological
model and describe numerical simulations used in our analysis. The
mass function of simulated halos span a mass range of seven orders of
magnitude. Density profiles of the most resolved dwarf halos from the
1 \mpch\ box are presented in \S 3.1. The concentration versus mass
diagram along with the analytical prediction by \citet{Bullock2001a}
is also shown there.  In \S 3.2 we compute the subhalo velocity
function for our most massive halos and show the almost self-similar
nature of the dark matter inside virialized systems, in agreement with
the results of \citet{Moore2001} and \citet{Delucia03}.  The
differential mass function of halos from the 1 \mpch\ box is
calculated in \S 4.1. A comparison is made with the prediction by
\citet{ST99}. In \S 4.2 we calculate the spin parameter distribution
of dwarf halos for the 1 \mpch\ box and compare it with one at $z = 0$
with the same cosmology. We discuss some of the results presented in
previous sections and present our concluding remarks in \S 5.

\section{Numerical Simulations}

We use a series of simulations of a low-density flat $\Lambda$CDM
cosmological model with the following parameters: $\Omega_0=0.3$,
$\Omega_{\Lambda}=0.7$, and $h=0.7$. All simulations are done with the
Adaptive Refinement Tree (ART) code \citep{KKK97}.  The ART code
achieves high spatial resolution by refining the base uniform grid in
all high-density regions with an automated refinement algorithm.

For our code a convergence study shows that density deviations less 
than 10\% are expected at radii larger than 4 times the formal force
resolution or the radius that contains 200 particles, whichever is
larger \citep{KKBP01}. This is recently confirmed in a detailed 
analysis of halo profiles by \citet{tkgk}.
With the aim of studying the structure of dwarf halos, we perform a
simulation of an 1 \mpch\ box on a side with $256^3$ particles.  One
of the motivations for choosing this small box is to reproduce the
same configuration as in \citet{Ricotti}.  The simulation has been
stopped at $z = 2.3 $ when the longest fluctuations in the box are
still well in the linear regime.  At redshift $z=3.3$ in the
simulation there are 4 large halos resolved with more than $200,000$
particles and with proper force resolution $14\pch$. For analysis we
also use many more smaller halos.

To complement our study we  use  three more
simulations. We use the simulation presented in \citet{KKBP01}, who
focused on a convergence study of the density structure of three Milky
Way-sized halos at low redshifts. At $z=3$ the three largest halos in
that simulation have a mass of about $10^{11}\ \msunh$ and are
resolved with $\sim 10^5$ particles.  For our analysis we also use 12
large halos with more than 5000 particles ($M > 6 \times 10^{9}\
\msunh$).

We also use a 80 \mpch\ box simulation, in which at redshift $z = 3$ we
identified 5 large halos resolved with more than $10^5$ particles.
At $z = 0$ the halos have become a cluster-sized halo of virial mass
$M_{vir}\approx 2\times 10^{14}\msunh$.  The high redshift halos are
used for our analysis.

Finally, we use
the 60 \mpch\ simulation with 256$^3$ particles described in
\cite{colin99,Klypin1999a}. The parameters of all the simulations are
presented in Table~\ref{tab:simu}. In column 1 we show the comoving
size of the computational box. The mass resolution is given in column
2, while the formal spatial resolution (size of a cell in the finest
refinement grid in comoving units) is shown in column 3. Column 4
gives the value of \sige\ (the rms of mass fluctuations estimated with
the top-hat window of radius $8 \mpch$). Finally, column 5 gives the
name of the simulation (which is also used for objects studied in this
simulation).

In our simulations the halos are identified by the
Bounded-Density-Maxima (BDM) algorithm
\citep{KlypinHoltzman,Klypin1999a}.  The BDM algorithm first finds
positions of local maxima in the density field.  These density maxima
are found using a top-hat filter with a ``search radius'' which
is several times larger than the force resolution. Once centers of
potential halos are found, the algorithm identifies halos around them
and removes particles which are not bound to those halos. This
procedure also detects subhalos of larger objects -- halos inside  halos
(for example, satellites of galaxies or galaxies in
clusters). Particles of a subhalo are bound to both the subhalo and to
the larger halo.

\begin{planotable}{crrrrcc}
\tablecolumns{7}
\tablewidth{0pc}
\tablecaption{Parameters of the most massive halos at $z\approx 3$}
\label{tab:halo}
\tablehead{\colhead{$M_{vir}$} & \colhead{$r_{vir}$} &\colhead{$N_{part}$} & \colhead{\Vmax} & 
\colhead{$r_s$}& \colhead{$\rho_0$}& \colhead{Halo name} \\
 (\msunh)& (\kpch)& & (\kms) & (proper \kpch)  & (proper \msun $h^2$ pc$^{-3}$) & }
\startdata
     $4.6 \times 10^9$ &    9.3 & 933636 &   48 &  2.0 & $5.0 \times 10^{-2}$ & A1 \\
     $1.4 \times 10^9$ &    6.6 & 285651 &   34 &  1.1 & $7.8 \times 10^{-2}$ & A2 \\
     $1.1 \times 10^9$ &    6.1 & 227275 &   28 &  1.1 & $6.6 \times 10^{-2}$ & A3 \\
     $1.0 \times 10^9$ &    6.0 & 208005 &   31 &  1.0 & $7.6 \times 10^{-2}$ & A4 \\
 $2.9 \times 10^{11}$ &  41.7 & 238083 & 190 &  7.1 & $5.8 \times 10^{-2}$ & B1 \\
 $1.2 \times 10^{11}$ &  30.8 &   95899 & 128 &12.2 & $9.2 \times 10^{-3}$ & B2 \\
 $1.1 \times 10^{11}$ &  29.9 &   86832 & 136 &  4.4 & $8.2 \times 10^{-2}$ & B3 \\
 $1.2 \times 10^{13}$ &140.0 & 288061 & 647 &26.9 & $4.8 \times 10^{-2}$ & C1 \\
 $6.9 \times 10^{12}$ &117.3 & 175882 & 497 &57.2 & $6.0 \times 10^{-3}$ & C2 \\
\enddata
\end{planotable}

\section{Halo structure}

\subsection{Density profiles and concentrations}

\begin{figure}[htb!]
\plotone{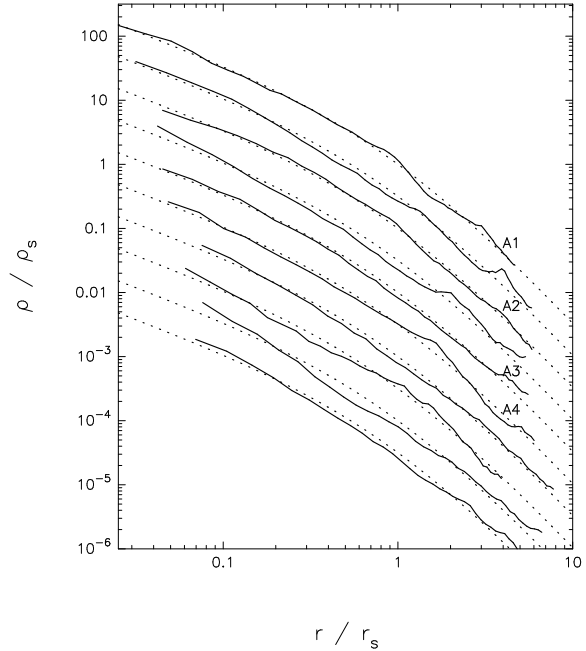}
\caption{Density profiles of the ten most massive halos of the 1
\mpch\ box in normalized units.  The  dotted curves iare the NFW
profile.  For clarity we shift down the profiles by $-0.5$ in
logarithm.  Halos are ordered by mass with the most massive halo being at
the top of the plot.  The density profiles are cut at the virial
radii.  The innermost bin contains more than 200 particles. No sign of
a shallow core is seen in any of the halos. However, most halos
have wiggles in the peripheral regions, which  are due to
residual substructure and ongoing merging. The labels A$n$, with $n=1,...,4$, 
represent the ``equilibrium'' halos used for subsequent analysis.}
\label{fig:tenhalos}
\end{figure}

We start our analysis of halo profiles by studying all most massive
halos (subhalos are not included). The profiles are not averaged over
time and halos are not selected to be quite or not to have large
merging events. Thus, some non-equilibrium features are expected.
Figure~\ref{fig:tenhalos} shows the density profiles of the ten most
massive halos in the $1\mpch$ box at $z=3.3$ plotted from the radii
which contain at least 200 particles to their virial radii.
The profiles are given in
normalized units: the radius is measured in units of
the scale radius $r_s$, the radius where the logarithmic
derivative of the NFW profile (see equation 1) is equal to $-2$,
whereas the density is measured in units of
a characteristic density $\rho_s = \rho_0/4$,
the NFW density evaluated at $r =r_s$.
These parameters are
taken from their corresponding NFW fits. The halos show a variety of
profiles. Some profiles have significant deviations from the NFW fit
(e.g., the second and the forth from the top) while others are fitted
by NFW reasonably well (e.g., the fifth halo).

The wiggles in the profiles seen
in the outskirts of most halos can be explained as substructure that
have not yet reached equilibrium with the rest of the halo or as the
result of a significant merger. For example, halos number 2 and 9
from top are in the process of a major merger. Both halos show a
density enhancement in the peripheral part (more pronounced in case
of halo 2). The almost power-law profile of halo 9 may be a
consequence of the way the merger has developed
\footnote{Out of the fifteen biggest halos in the 1 \mpch\ box, the
merger companion of halo 9 has the most peculiar density profile. It
has the highest $r_s$ value even though it is the second {\it least
massive} halo on the list.}.

No matter how large or small are the deviations, the profiles do not
show a sign of flat cores in the central part of dwarf halos.  In
other words, steep central cusps of halos in our simulations
contradict the results of \citet{Ricotti}, who finds that central slopes
of halos in his 1 \mpch\ simulation are very shallow: $\rho \propto
r^{\alpha}$, $\alpha \approx -0.2$. 

There are some differences in the way how we and \citet{Ricotti} approximate
profiles. Ricotti does the approximations at much earlier time
$z \sim 10$ when halos are much smaller and are less
resolved. Instead of making two-parameter fits as we do, Ricotti uses
a very simple approach: he uses ``location and value of the maximum
circular velocity''. When we apply the same prescription to our high
resolution profiles at $z=3$, we get very noisy results with fits
going either above or below actual profiles. Yet, we did not get
sytematic effects. In Figure 2 we show the density profiles of the
6 most massive halos at $z = 10.5$ fitted with NFW profiles using
two free parameters. The NFW fits approximate reasonable well the
density profiles. Even without any fits it is clear that none of 
the profiles are even close to the shallow profiles suggested by 
Ricotti.

\begin{figure}[htb!]
\epsscale{1.10} 
\plotone{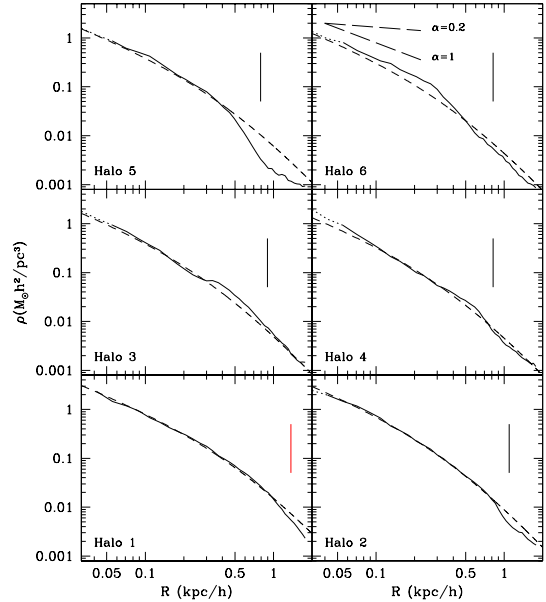}
\caption{Density profiles of the most massive halos at $z=10.5$ 
(full curves). Dashed curves show NFW
fits. Long-dashed lines in the plot of Halo6 show power-laws with
slopes $\alpha=1$(NFW) and $\alpha=0.2$(Ricotti). In spite of large
fluctuations at large (virial) radii, none of the halos show slopes
suggested by \citet{Ricotti}.}
\end{figure}

In the subsequent analysis we mostly focus on equilibrium halos, which
are fitted by the NFW profile.  Table~\ref{tab:halo} presents
parameters of the most massive halos identified in the simulations A
-- C at redshift $z \sim 3$.  In column 1 we show the mass \Mvir\
within the virial radius (column 2). Here and below we present radii
and densities in proper units.  The virial radius, \rvir, is
defined as the radius where the average
halo density is $\delta$ times the background density according to the
spherical top-hat model. Here $\delta$ is a number that depends on
epoch and cosmological parameters (\ome,\omel); for a flat \LCDM\
model, $\delta \sim 180$ at $z = 3$.  In column 3 the corresponding
number of particles is given. The maximum circular velocity $ v_{\rm
max}$ $=\left( GM(<r)/r \right)^{1/2}_{\rm max},$ where $G$ is the
gravitational constant and $M(<r)$ is the mass within the radius $r$,
is located in column 4. In column 5 and 6 we present the NFW
parameters $r_s$ and $\rho_0$ in proper units.  The name of the halos
are given in the last column.

We impose the following ``equilibrium'' criteria to select a halo for
further analysis: (1) The halo should not be within the virial radius
of a more massive halo; i.e, the halo is not a subhalo. (2) The halo
should not have a subhalo with a mass larger than one tenth the mass of
the halo.  Finally, (3) we also use a goodness of fit criterion to
reject halos with $D_f \ge 0.004$, where the accuracy of the fit $D_f$
is defined as follows.

We compute the density, $\rho(r)$, in spherical shells whose radii
increase as the square of the bin number, $n$. Under this scheme, the
logarithmic radial width, $\Delta \log r$, is not constant but decreases 
with $n$, and for high values of $n$ $\Delta \log r$ goes as $2/n$.
This  binning  improves  the often used 
constant logarithmic binning.
We use the NFW profile, $\rho_{\nfw}$, to fit our halo
density profiles
\begin{equation}
\rho_{\nfw} (r) = \frac{\rho_0}{r/r_s(1 + r/r_s)^2}.
\end{equation}
The best fit and thus parameters $\rho_0$ and $r_s$ are obtained 
by minimizing $\chi^2$,
\begin{equation}
\chi^2 (\rho_0,r_s) = \sum_{n=1}^N \left[ \frac{\log \rho(r_n) 
- \log \rho_{\nfw}(r_n)}{\sigma_n} \right]^2.
\end{equation}
We  define $\omega_n \equiv 1/\sigma_n^2$ to be 
\begin{equation}
\omega_n = \frac{\Delta r}{r_n} = \frac{2n + 7}{(n + 3)^2}
\end{equation}
 to compensate for the increasing density of bins with radius.  In
equation (3), $\Delta r$ is the radial width for our binning scheme
and $n$ is the number of the bin corresponding to radius $r_n$.
Notice that by definition $\omega$ is constant for a binning constant
in $\log(r)$. We estimate the accuracy of the fit with the parameter
$D \equiv \chi^2 /F$, where $F = \sum \omega_n$ and the sum runs over
the number of bins $N$.  This parameter reduces to $\chi^2/N$ (with
$\sigma_n = 1$) when $\omega$ is constant and it is a better
estimation of the goodness of the fit than, for example, a maximum
deviation because average out possible large fluctuations in some
bins. We  set $D = D_f = 0.004$ as our criterion for the quality
of the fit. Halos with $D_f < 0.004$ that satisfy at the same time our
minimum requirements for equilibrium (criteria 1 and 2) comprise
65\% of the sample. In fact, in the distribution of $D$ there is a
strong drop in the number of halos above $D_f$. It seems that  $D_f$
separates relaxed from  unrelaxed halos.

\begin{figure}[htb!]
\plotone{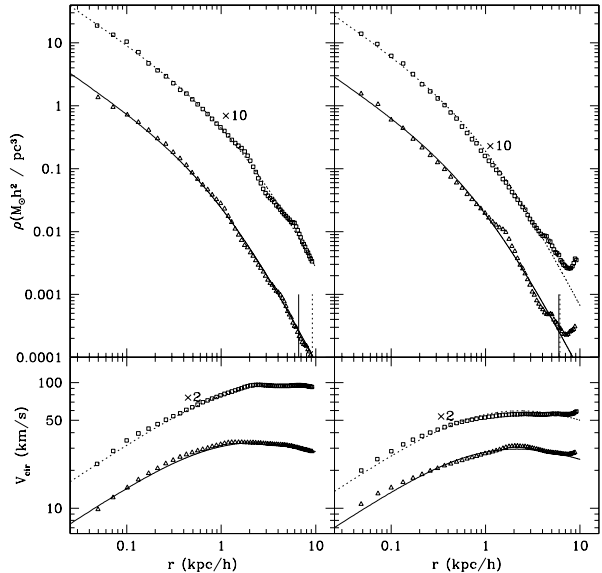}
\caption{Density and circular velocity profiles of the four most
massive halos at $z = 3.3$ that satisfy our criteria of
``equilibrium'' halos.  The densities of the first (A1, left panel)
and third (A3, right panel) halos have been multiplied by ten to
prevent overlapping. The corresponding circular velocities (lower
panel) were multiplied by a factor of two. The innermost bin is chosen
to contain at least 200 particles. Vertical lines at the bottom of the
density panels mark the corresponding virial radii (A1, A3 - dotted
lines, A2, A4 solid lines)}
\end{figure}

Figure 3 shows the density and circular velocity, \Vc, profiles for
the four most massive halos of the 1 \mpch\ box that satisfy the
criteria of ``well-behaved'' or ``equilibrium'' halos (A1 --
A4). Dotted and solid lines are the NFW best fits. NFW circular
velocity profiles are computed using $(\rho_0 ,r_s)$ parameters
obtained from the density fits.  The NFW profile describes quite well
both the density and the circular velocity profiles of these halos.
At redshift $z \sim 3$ the majority -- about 65\% -- of dwarf halos
with more than 5000 particles are in ``equilibrium''.

\begin{figure}[htb!]
\epsscale{1.20} 
\plotone{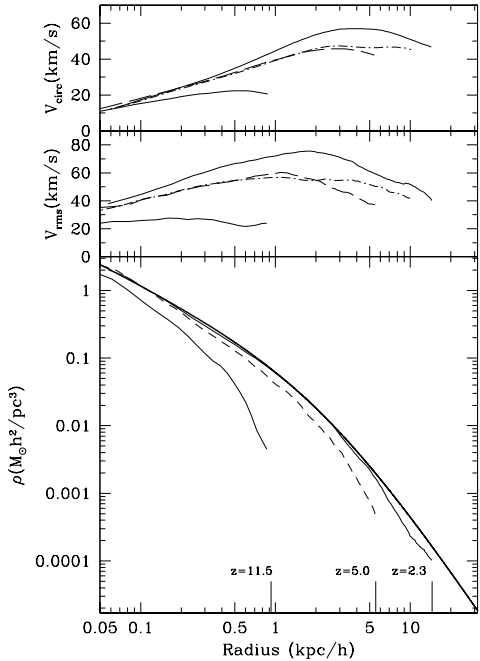}
\caption{Circular velocity, 3D velocity dispersion, and density profiles
of the most massive halo A1 at different redshifs. Dot dashed curves
on the top two panels are for $z=3.3$. For clarity we do not show the
density profile for this redshift.  Vertical lines in the bottom panel
mark the virial radii. Most of the changes happen at outer radii of
the halo while there is very little evolution in the center. The thick
full curve in the bottom panel shows the NFW fit for $z=2.3$.}
\label{fig:haloevol}
\end{figure}

Figure~\ref{fig:haloevol} shows the inner structure of the most
massive halo A1 (density, circular velocity and 3D velocity dispersion
profiles) for epochs from $z = 11.5$ to $z = 2.3$. The virial mass of
the halo increases 71 times during the period.  Yet, the inner density
profile (radii less than $\approx 100$~pc) changes very little. Thus,
the cusp of the halo was already in place at very high redshift and
since then shows little evolution. The central 0.5~kpc region shows
little evolution since redshift $z=5$.  This does not mean that
particles, which at redshift $z\approx 10$ are in the core stay there
at later times. Comparison of the rms velocities (middle panel) with
the circular velocities (top panel) clearly indicates that particles
in the core move with random velocities that significantly exceed the
circular velocity. This implies that the core itself is not self-bound
and a large fraction of particles simply passes through the cusp.
This is very important for understanding the evolution of halo 
concentration.

\begin{figure}[htb!]
\epsscale{1.0} 
\plotone{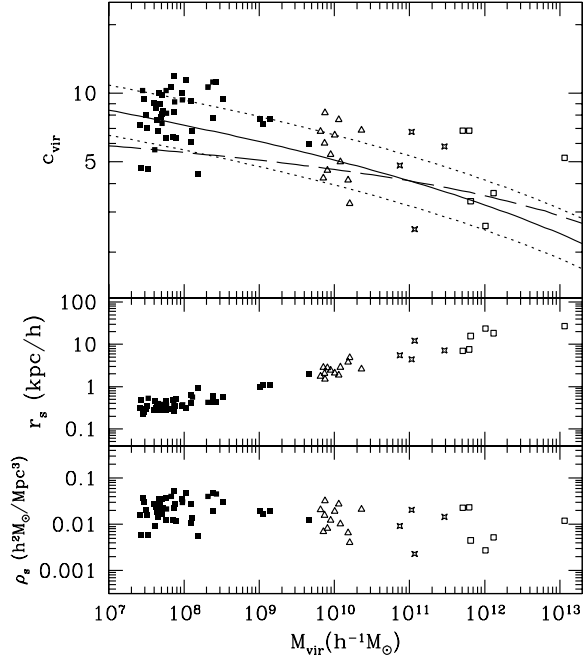}
\caption{ The dependence of concentration ($c_{vir}$, top panel), the
scale radius ($r_s$, middle panel) and $\rho_s = \rho_0/4$
($\rho_s$, bottom panel) on the halo virial mass. Solid squares are
halos in the 1 \mpch\ box at $z = 3.3$. The rest of the points are for
halos in the 25 \mpch\ simulation (open triangles and four-point
stars) and the 80 \mpch\ simulation (open squares) at $z = 3$.  The
solid and dashed curves represent the predictions of the models by \citet{Bullock2001a}
and \citet{ENS2001}, respectively, while 
the dotted lines are the 1$\sigma$ scatter of halos from the 1 \mpch\
box, $\Delta(\log c_{vir}) = 0.11$.}
\end{figure}

For analysis of halo concentrations we make fits for halos with more
than 5000 particles in all our simulations. At $z = 3.3$ in the 1
\mpch\ box simulation there are 72 of these halos of which 40 satisfy
the equilibrium criteria. We also include seven more halos that do not
satisfy the second criterion (have large companion), but are still
well fitted by the NFW profile.  Halos selected in the other two
simulations also have more than 5000 particles each and the deviations
of the measured density profile from the NFW fit are less than 30\%;
$\left| \rho(r_i) - \rho_{\nfw}(r_i) \right| / \rho_{\nfw}(r_i) <
0.3$, where $r_i$ goes from the first radial bin to \rvir. Halos that
are too disturbed hardly satisfy this latter ``equilibrium''
requirement \citep{Jing}.  Figure 5 shows concentrations $c_{vir}
\equiv r_{vir}/r_s$, proper scale radii, $r_s$, and proper densities
at the scale radii $\rho_s = \rho_0/4$ for halos of different virial
masses \Mvir.  The solid and dashed curves are the predictions of 
the concentration at $z = 3.0$ based on the models of \citet{Bullock2001a} 
and \citet{ENS2001}, respectively, for the
appropriate cosmological model with $\sige = 0.9$.  The points from
the 1 \mpch\ box are slightly scaled up to take into account the
different normalization ($\sige = 0.75$) and the slightly different
identification redshift ($z = 3.3$). For this latter, we use the $1
+z$ dependence of $c_{vir}$ while for the former we multiply $c_{vir}$
by $0.9/0.75$, which is a good approximation because dwarf halos form
very early when the growth factor is proportional to
$a$.\footnote{Bullock et al. (2001a) define the epoch of collapse as
the epoch at which the non-linear mass, $M_*$, equals a fixed fraction
$F$ of the halo mass, \Mvir. For the \LCDM\ model, Bullock et
al. suggest to use $F = 0.01$.} We use the halos from the 1 \mpch\ box
to compute the standard deviation of the $\log(c)$ distribution and
find $\Delta(\log c_{vir}) = 0.11$, which is a factor 1.6 lower than
that quoted by \citet{Bullock2001a}.  If we use the whole sample of
halos only under the restriction that they must have more than 5000
particles, the scatter increases to 0.14.  So, at least part if not
all (see \S 5), of our smaller scatter is due to the fact that we are
using relaxed halos.

In summary, dwarf halos seem to be in many respects a down-scaled copy
of more massive halos. For example, they all have cusps regardless of
their equilibrium status. Interestingly, they have a median
concentration that agree with the model of \citet{Bullock2001a}. We
extended the $c_{vir}$ versus \Mvir\ diagram to high masses and showed
how the Bullock et al. model continue to predict the correct concentration.

\subsection{Substructure mass function}

The cumulative velocity function of subhalos within halos A1 (open
circles), B1 (open triangles), and C1 (solid circles) are compared in
Figure 6. The peak velocities of subhalos are measured in units of the
virial velocity, defined as the circular velocity at the virial radius,
of their parent halo.  The mass of the halos covered almost four orders
of magnitude yet the subhalo velocity distributions are similar
\citep{Moore2001, Delucia03}. This function can be approximated by a
power-law with power index $-2.75$ \citep{Klypin1999b} (see dot-dashed
line in Figure 6).

\begin{figure}[htb!]
\epsscale{1.0} 
\plotone{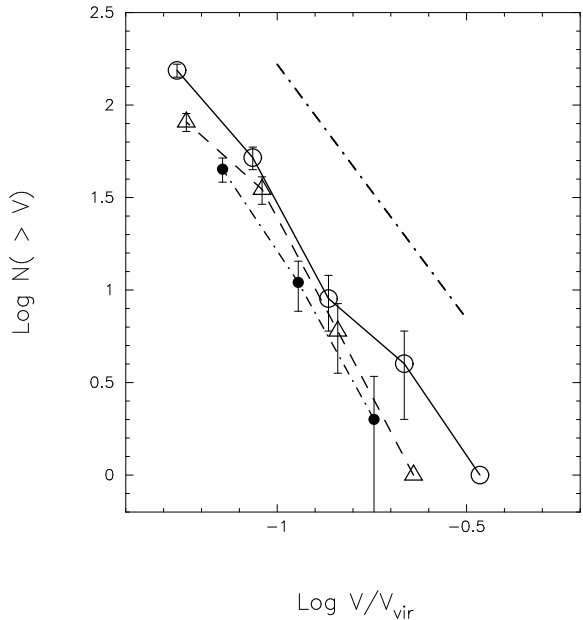}
\caption{
The cumulative number  of
subhalos with given circular velocity for halos A1 (open circles), B1 (open triangles), and C1
(solid circles). At $z \sim 3$ each halo has more than $2 \times 10^5$
particles and is resolved down to radius 0.01\rvir (the first radial
bin with $> 200$ particles). The subhalo maximum circular velocities
are in units of the virial velocity of the parent halo. The dot-dashed
line shows the power law with the slope -2.75 found in previous
simulations.}
\end{figure}

\section{Statistics of halos}

\subsection{Halo mass function}

The mass function of halos $N(M)$ is defined so that $N(M) dM dV$
represents the number of halos with masses between $M$ and $M + dM$ in
the volume $dV$. This is a differential mass function.
In order to find isolated halos we use the friends-of-friends group
finding algorithm \citep{Davis1985} with a linking length of 0.2 (or in
physical units 0.78 \kpch). Halos are identified as the set of
particles mutually linked. The mass of the halo is simply the sum of
its constituent dark matter particles. Tests \citep[\eg][]{governato99}
indicate that at least 30 particles are needed to take into account
properly the effect of mass resolution.  In this paper we analyze only
halos with more than 40 particles.

We construct the dwarf halo mass function by grouping the halos
according to their mass in bins of $\Delta \log (M) = 0.1$.  In order
to avoid too large statistical fluctuations, we do not use the few
most massive halos. Error bars for data in individual bins are
estimated using Poisson statistics.

\begin{figure}[htb!]
\plotone{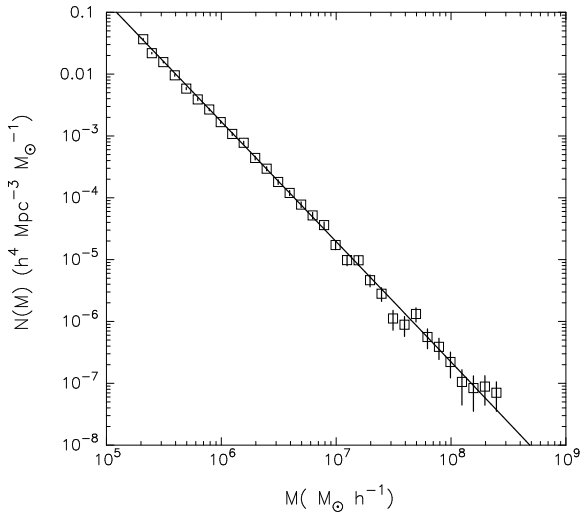}
\caption{The halo mass function from the 1 \mpch\ box (open squares).
Error bars are computed assuming the Poisson statistics. The solid line
is the analytical prediction from the Sheth-Tormen formalism.}
\end{figure}

Figure 7 shows the mass function (open squares) of our dwarf
halos from the 1 \mpch\ box. The solid line is the analytical prediction
of the mass function from the Sheth-Tormen formalism.
Notice the excellent agreement between the prediction of the
Sheth-Tormen model and our numerical results. This model has been
tested under diverse conditions: redshifts, cosmological parameters,
halo masses ($> 10^{10} \msunh$), etc., and have passed all tests (but
see Reed et al. 2003).  Here we see that it successfully predicts the
mass function of dwarf halos, albeit at high redshift. In summary, the
Sheth-Tormen formalism can be safely applied in different cosmological
models and for about ten orders of magnitude in mass. With the
corresponding correction factors the Sheth-Tormen formalism can be
successfully applied also to predict the number of dwarfs in voids
\citep{gottloeber03}.

\subsection{Spin parameter distribution}

To study the distribution of the spin parameter, $p(\lambda)$, we use
halos which do not reside inside a larger halo and have more than 500
particles.  The spin parameter $\lambda$ is defined by
\begin{equation}
\lambda \equiv \frac{J \left| E \right|^{1/2}}{G M^{5/2}},
\end{equation}
where $J$ is the magnitude of the angular momentum, $E$ is the 
total energy, and $M$ is the mass of the halo. We assume
that halos are in virial equilibrium and thus $E = - K$, where
$K$ is the kinetic energy inside \rvir. N-body simulations
 have shown that $p(\lambda)$ is well described by a lognormal
distribution (see \cite{maya2002} and references therein)
\begin{equation}
p(\lambda) d\lambda = \frac{1}{\sigma_{\lambda} \sqrt{2 \pi}} 
\exp \left[-\frac{\ln^2 \left( \lambda/\lambda_0 \right)}
{2 \sigma_{\lambda}}\right] \frac{d \lambda}{\lambda}.
\end{equation}
Figure 8 compares the probability distributions of the 
spin parameter in simulations A and D (Table~\ref{tab:simu}) and
shows the lognormal fits. 

\begin{figure}[htb!]
\plotone{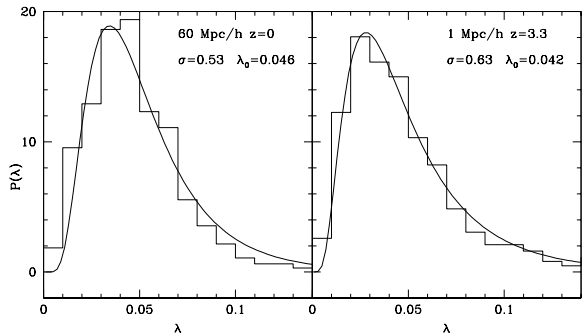}
\caption{Spin parameter distribution (histogram) for
halos with more than 500 particles drawn from a 60 \mpch\ box 
simulation at $z = 0$ (left panel) and from our 1 \mpch\ box
at $z = 3.3$ (right panel). Curves on each panel are lognormal
best-fits to the data. The parameters of the fit 
$(\sigma_{\lambda},\lambda_0)$ are also shown in panels.}
\end{figure}

Figure 8 shows that the distribution of the spin parameter
of dwarf halos also follows a lognormal law. In
fact, it turns out that dwarf halos in simulations A are better described
by equation (5) than the more massive halos in the
simulation D: their distribution has a lower $\chi^2$
for the same number of bins. According to the values of the
parameters of the fit, it would appear that the
distributions are different, but within the errors they 
are similar. What happens is that $p(\lambda)$ is very
sensitive to $(\sigma_{\lambda},\lambda_0)$. In general,
our results
confirm the known independence of the spin parameter on
mass \citep{lk99} and epoch \citep[\eg][]{maya2002}.

\section{Discussion and conclusions}

We find that the density profiles of relaxed dwarf halos
can be well fitted by the NFW profile, which has an inner density
slope $\alpha = -1$. This result is in conflict with results of
\cite{Ricotti}, who finds $\alpha\approx -0.5$ for dwarf halos in his
1 \mpch\ box simulation. In a attempt to find the cause of our
discrepancy, we studied density profiles at the same redshift as Ricotti
did ($z = 10.5$) and reproduced Ricotti's fitting procedure. The
conclusion is that even without any fits none of 
the profiles are close to the shallow profiles suggested by 
Ricotti. On the other hand, \citet{Navarro03}, in a recent study of 
mass profiles of \LCDM\ halos at $z = 0$ that span five decades 
in halo mass, show that inner density slopes of dwarf halos are steeper than
$-1$ all the way down to the resolution limit. From their Tables
2 and 3 we find dwarf halo concentrations from $c_{vir} = 16.4$
for halo D1 to $c_{vir} = 22.0$ for halo D4. These concentrations
are below to our extrapolated value of the 
median concentration we find for our dwarf halos, but once the
mass dependence of the concentration is considered they turn out
to be similar. Our results thus agree with the study of Navarro et al.
on dwarf halos not only in the inner shape of density profiles
but also on concentrations.

It is impressive to see how well the model of \citet{Bullock2001a}
reproduces over five orders of magnitude the measured concentration
as a function of halo mass. At the same time, Bullock et al. 
overpredict the deviations from the mean value. We compute the median
and standard deviation of the $c_{vir}$ distribution of our sample of
relaxed dwarf halos and find a spread which disagrees with that
reported in Bullock et al. by about a factor 1.6 in $\Delta \log c_{vir}$.
It appears that the printed version of Bullock et al. has a
typo and the actual value for the spread of concentration in
Bullock et al. should have been $\Delta \log c_{vir} = 0.14$ (Bullock,
private communication). This is still larger than what we found:
$\Delta \log c_{vir} = 0.11$.  Our 1$\sigma$ scatter is close to $\Delta \log c_{vir} = 0.10$
measured by \cite{klypin03}. We
found that, except at the high-mass end, Eke's et al. model lie below
Bullock's et al. For instance, at $\Mvir = 10^7 \msunh$ Bullock's et al. model
predicts $c_{vir} = 8.8$ as compared with 5.5 of Eke's et al. model.

If we extrapolate the concentration of our dwarf halos to the present 
using the 1+$z$ growth law of the Bullock et al. model, the median 
would be about 30. This is higher; for example, than the value
$c_{200} \sim 14$ (or $c_{vir} \sim 18$, no mass loss case) 
used by Hayashi et al. (2003). A shift up in $c_{200}$ in 
Hayashi et al. fitting procedure would reduce the lower limit in \Vmax; 
that is, a lower \Vmax\ would be required to match
the observed stellar velocity dispersion of Carina and Draco. In
any case, it is not clear to what extent Hayashi et al. results can
be applied to the observed satellites when substructure is
constrained to evolve in a static potential. 
Stoehr et al. (2002), on the other hand, are
able to explain the observed central velocity dispersions of the
dSph satellites of the Milky Way for the 20 most massive subhalos
of a Milky Way-sized galaxy halo. Subhalos are resolved with 
hundreds or at most with a few thousands of particles. These 
still small numbers of particles
are probably not enough to draw yet firm conclusions. In the light
of our results, it seems that subhalos resolved with
many more particles than subhalos in present-day 
cosmological simulations
will be more concentrated. However, it is not clear how
this effect will affect the conclusion
that observed dSph galaxies inhabit the largest subhalos.

In our analysis of the structure of dwarf halos, we also
compute the abundance of subhalos. We measure the
amount of substructure in halos A1 (dwarf halo), B1 
(Magellanic Cloud-sized halo) and C1 (group-size halo)
through the normalized subhalo \Vmax\ function (\Vmax\ is 
measured in units of the virial velocity of the parent halo)
and find a mass-scale independent result: the curves
are similar and well approximated by the power-law $n(>V) 
\propto V^{-2.75}$ \citep{Klypin1999b}. This is in 
agreement with the paper by \citet{Moore2001} (see also
\citet{Delucia03}).

We compute the dwarf halo mass function and find an excellent
agreement with the model of \citet{ST99}. The model has been tested at
different redshifts, for a variety of cosmologies, and for halo masses
above $\sim 10^{10} \msunh$ \citep{ST99, jenkins01, reed03}.  It
provides an excellent match to the data. In this paper, we tested it
at much lower masses, $10^5 -10^8 \msunh$, and found also an
excellent agreement with the data. It is pleasant to know that this
approximation is able to match data for ten orders of magnitude in
mass.

The spin parameter distribution $p(\lambda)$ is measured for our
sample of dwarf halos (halos with a virial mass between $10^7$ and
$10^9$ \msunh). It is well fitted by a lognormal distribution with
$\lambda_0 = 0.042$ and $\sigma_{\lambda} = 0.63$. These parameters
roughly agree with those estimated by \citet{maya2002}.  Our results
corroborate the known independence of the spin parameter on mass
\citep{lk99} and epoch \citep[\eg][]{maya2002}.

Acknowledgement: P.C. acknowledge support by CONACyT grant 36584-E,
S.G. and A.K. by NSF/DAAD, S.G. and P.C. by DFG/CONACyT.  Computer
simulations presented in this paper were done at the
Leibnizrechenzentrum (LRZ) in Munich and at the National Energy
Research Scientific Computing Center (NERSC).
We thank J.~Bullock and V.~Avila-Reese for helpful comments and discussions. 
We acknowledge the anonymous referee whose helpful comments and
suggestions improved some aspects of this paper.


\begin{thebibliography}{DUM}
\bibitem[Benson et al. (2002)]{Benson2002}
        Benson, A.J., Frenk, C.S., Lacey, C.G., Baugh, C.M., \& Cole, S.
        2002, MNRAS, 333, 177
\bibitem[Bond et al. (1991)]{Bond1991}
        Bond, J.R., Cole, Efstathiou, G., \& Kaiser, N. 1991, apJ, 379, 440
\bibitem[Bullock, Kravtsov, \& Weinberg (2000)]{BKW2000}
        Bullock, J.S., Kravtsov, A.V., \& Weinberg, D.H. 2000, ApJ, 539, 517
\bibitem[Bullock et al. (2001a)]{Bullock2001a}
        Bullock, J.S., Kolatt, T.S., Sigad, Y., Somerville, R.S., Kravtsov
        A.V., Klypin, A.A., Primack, J.R., \& Dekel, A. 2001a, MNRAS, 321, 559
\bibitem[Bullock et al. (2001b)]{Bullock2001b}
        Bullock, J.S., Dekel, A., Kolatt, T.S., Kravtsov, A.V., Klypin, A.A.,
        Porciani, C., \& Primack, J.R. 2001, ApJ, 555, 240
\bibitem[Burkert (1995)]{Burkert1995}
        Burkert, A. 1995, ApJ, 447, L25
\bibitem[Carignan \& Beauliu (1989)]{CB1989}
        Carignan, C., \& Beauliu, S. 1989, ApJ, 347, 760
\bibitem[Carignan \& Freeman (1988)]{CF1988}
        Carignan, C., \& Freeman, K.C. 1988, ApJ, 332, L33
\bibitem[de Blok, Bosma, \& McGaugh (2003)]{deBlok}
        de Blok, W.J.G., Bosma, A., \& McGaugh, S. 2003, MNRAS, 340, 657
\bibitem[Cen et al. (2004)]{Cen04}
        Cen, R., Dong, F., Bode, P., \& Ostriker, J.P. 
        ApJ submitted (astro-ph/0403352)
\bibitem[Col\'in et al. (1999)]{colin99}
        Col\'in, P., Klypin, A.A., Kravtsov, A.V., \& A.M. Khokhlov. 1999,
        ApJ, 523, 32
\bibitem[De Lucia et al. (2003)]{Delucia03}
        De Lucia, G., Kauffmann, G., Springel, V., \& White, S.D.M. 
        MNRAS submitted (astro-ph/0306205)
\bibitem[Davis et al. (1985)]{Davis1985}
        Davis, M., Efstathiou, G., Frenk, C.S., \& White, S.D.M.
        1985, ApJ, 292, 371
\bibitem[Dubinski \& Carlberg (1991)]{DC1991}
        Dubinski, J., \& Carlberg, R. 1991, ApJ, 378, 496
\bibitem[Eke, Cole, \& Frenk (1996)]{Eke1996}
        Eke, V.R., Cole, S., Frenk, C.S. 1996, MNRAS, 282, 263
\bibitem[Eke, Navarro, \& Steinmetz (2001)]{ENS2001}
        Eke, V.R., Navarro, J.F., \& Steinmetz, M. 2001, ApJ, 554, 114
\bibitem[Flores \& Primack (1994)]{FP1994}
        Flores, R.A., \& Primack, J.R. 1994, ApJ, 427, L1
\bibitem[Gottl\"ober et al. (2003)]{gottloeber03}
        Gottl\"ober, S., {\L}okas, E., Klypin, A.A., Hoffman Y. 2003,
        MNRAS, in press 
\bibitem[Governato et al. (1999)]{governato99}
        Governato, F., Babul, A., Quinn,  T., Tozzi, P., Baugh,  C.M., 
        Katz, N., \& Lake, G. 1999, MNRAS, 307, 949 
\bibitem[Hayashi et al. (2003)]{Hayashi2003}
        Hayashi, E., Navarro, J.F., Taylor, J.E., Stadel, J., \& Quinn, T.
        2003, ApJ, 584, 541
\bibitem[Jenkins et al. (2001)]{jenkins01}
        Jenkins, A., Frenk, C.S., White, S.D.M., Colberg, J.M., Cole, S.,
        Evrard, A.E., Couchman, H.M.P., \& Yoshida, N. 2001, MNRAS, 321, 372
\bibitem[Jing (2000)]{Jing}
        Jing, Y.P. 2000, ApJ, 535, 30
\bibitem[Klypin \& Holtzman (1997)]{KlypinHoltzman} Klypin, A.A., Holtzman, J., 1997, 
        http://xxx.lanl.gov/pdf/astro-ph/9712217
\bibitem[Klypin et al. (1999a)]{Klypin1999a}
        Klypin, A.A.,  Gottl\"ober, S., Kravtsov, A.V., \& Khokhlov, A.M
        1999, ApJ, 516, 530
\bibitem[Klypin et al. (1999b)]{Klypin1999b}
        Klypin, A.A., Kravtsov, A.V., Valenzuela, O., \& Prada, F.
        1999, ApJ, 522, 82
\bibitem[Klypin et al. (2001)]{KKBP01} Klypin, A.A., Kravtsov, A.V.,
        Bullock, J.S., \& Primack, J.R. 2001, ApJ, 554, 903 
\bibitem[Klypin et al. (2003)]{klypin03}
        Klypin, A.A., Macci\'{o}, A.V., Mainini, R., \& Bonometto, S.A.
        2003, ApJ submitted (astro-ph/0303304)
\bibitem[Kravtsov et al.(1997)]{KKK97}
        Kravtsov, A.V., Klypin, A.A., \& Khokhlov, A.M., 1997, ApJS, 111, 73
\bibitem[Lemson \& Kauffmann (1999)]{lk99}
        Lemson, G., \& Kauffmann, G. 1999, MNRAS, 302, 111
\bibitem[\L okas (2002)]{lokas02}
        \L okas, E.L. 2002, MNRAS, 333, 697
\bibitem[Mateo (1998)]{mateo98}
        Mateo, M. 1998, ARA\&A, 36, 435
\bibitem[Moore (1994)]{Moore94}
        Moore, B. 1994, Nature, 370, 629
\bibitem[Moore et al. (1999)]{Moore1999a}
        Moore, B., Quinn, T., Governato, F., Stadel, J., \& Lake, G. 1999a, MNRAS,
        310, 1147
\bibitem[Moore et al. (1999)]{Moore1999b}
        Moore, B., Ghigna, S., Governato, F., Lake, G., Quinn, T., Stadel,
        J., Tozzi, P. 1999b, ApJ, 524, 19
\bibitem[Moore et al. (2001)]{Moore2001}
        Moore, B., Calc\'aneo-Rold\'an, C., Stadel, J., Quinn, T., Lake, G.,
        Ghigna, S., \& Governato, F. 2001, PhRvD, 64, 063508
\bibitem[Navarro, Frenk, \& White (1995)]{NFW95}
        Navarro, J.F., Frenk, C.S., \& White, S.D.M. 1995, MNRAS, 275, 720
\bibitem[Navarro, Frenk, \& White (1996)]{NFW96}
        Navarro, J.F., Frenk, C.S., \& White, S.D.M. 1996, ApJ, 462, 563
\bibitem[Navarro, Frenk, \& White (1997)]{NFW97}
        Navarro, J.F., Frenk, C.S., \& White, S.D.M. 1997, ApJ, 490, 493
\bibitem[Navarro et al. (2003)]{Navarro03}
	Navarro, J.F., Hayashi, E., Power, C., Jenkins, A.R., Frenk, C.S.,
        White, S.D.M., Springel, V., Stadel, J., \& Quinn, T.R. 2003, 
        MNRAS submitted (astro-ph/0311231)
\bibitem[Ricotti (2002)]{Ricotti}
        Ricotti, M. 2003, MNRAS, 344, 1237
\bibitem[Peebles (1980)]{Peebles80}
        Peebles, P.J.E. 1980, The Large Scale Structure of the Universe
        (Princeton: Princeton Univ. Press)
\bibitem[Reed et al. (2003)]{reed03}
        Reed, D., Gardner, J., Quinn, T., Stadel, J., Fardal, M., 
        Lake, G., \& Governato, F. 2003, MNRAS submitted (astro-ph/0301270)
\bibitem[Rhee, Klypin, \& Valenzuela (2003)]{Rhee2003}
        Rhee,G., Klypin, A., \& Valenzuela, O. 2003, in preparation
\bibitem[Sheth \& Tormen (1999)]{ST99}
        Sheth, R., \& Tormen, G. 1999, MNRAS, 308, 119
\bibitem[Somerville (2002)]{Somerville2002}
        Somerville, R. 2002, ApJ, 572, L23
\bibitem[Stoehr et al. (2002)]{Stoehr02}
        Stoehr, F., White, S.D.M., Tormen, G., \& Springel, V. 2002,
        MNRAS, 335, L84
\bibitem[Swaters et al. (2003)]{Swaters2003}
        Swaters, R.A., Madore, B.F., van den Bosch, F.C., \& Balcells, M.
        2003, ApJ, 583, 732
\bibitem[Tasitsiomi et al.(2003)]{tkgk}
	Tasitsiomi, A., Kravtsov, A.V., Gottl\"ober, S., \& Klypin, A.A.
	2003, ApJ submitted (astro-ph/0311062)
\bibitem[van den Bosch(2002)]{Bosch2002} van den Bosch, F.~C.\ 
2002, \mnras, 331, 98 
\bibitem[Vitvitska et al. (2002)]{maya2002}
        Vitvitska, M., Klypin, A.A., Kravtsov, A.V., Wechsler, R.H.,
        Primack, J.R., \& J.S. Bullock. 2002, ApJ, 581, 799
\bibitem[Weldrake, de Blok, \& Walter (2003)]{WdBW2003}
        Weldrake, D.T.F., de Blok, W.J.G., \& Walter, F. 2003, MNRAS, 340, 12
\end{thebibliography}
\end{document}